\documentstyle[11pt,epsf]{article}

\newcommand{\del}{\partial}
\newcommand{\beq}{\begin{eqnarray}}
\newcommand{\eeq}{\end{eqnarray}}
\newcommand{\be}{\begin{eqnarray*}}
\newcommand{\ee}{\end{eqnarray*}}

\newcommand{\ra}{\rightarrow}
\newcommand{\e}{\epsilon}

\newcommand{\ex}[1]{\langle\,#1\,\rangle}

\begin{document}

\centerline{\Large\bf {Radiative Corrections to the Casimir Energy}}
\vskip 10mm
\centerline{Xinwei Kong and Finn Ravndal}
\bigskip
\centerline{\it Institute of Physics}
\centerline{\it University of Oslo}
\centerline{\it N-0316 Oslo, Norway}

\bigskip
{\bf Abstract:} {\small The lowest radiative correction to the Casimir energy density between two
parallel plates is calculated using effective field theory. Since the correlators of the
electromagnetic field diverge near the plates, the regularized energy density is also divergent.
However, the regularized integral of the energy density is finite and varies with the plate
separation $L$ as $1/L^7$. This apparently paradoxical situation is analyzed in an equivalent, but
more transparent theory of a massless scalar field in 1+1 dimensions confined to a line element of
length $L$ and satisfying Dirichlet boundary conditions.
  
PACS numbers: 03.70.+k; 11.10; 32.80.Wr}

The remarkable prediction of quantum electrodynamics (QED) by Casimir\cite{Casimir} that
two parallel metallic plates in vacuum and with separation $L$ should be attracted by a force per
unit area equal to $\pi^2/240L^4$, has very recently been verified experimentally by
Lamoreaux\cite{exp} with much higher precision than before\cite{Sparnaay}. In fact, the experiment
is so accurate that with further improvements it can hopefully also demonstrate the corrections due to
finite-temperature effects\cite{Mehra}\cite{Brown}\cite{Schwinger}\cite{TR}. It would then be a
new macroscopic system where the intricacies of quantum field theory can be studied.

Despite an enormous literature on the subject (see for example the review article\cite{review}
and a recent textbook\cite{Milonni}) very little has been done to investigate the
radiative corrections to the this effect based on QED which is basically an interacting theory.
It is still not clear at which order in the fine structure
constant $\alpha$ the first correction will appear and how it will vary with the plate separation.
One of the first calculations were done by Bordag, Robaschik and Wieczorek\cite{Bordag} who found a
correction proportional to $\alpha/L^5$ in lowest order QED perturbation theory using a photon
propagator which is appropriately modified due to the plate confinement. Using instead periodic
boundary conditions, Xue found that the corresponding radiative corrections were
exponentially small\cite{Xue}.

Instead of using QED, we will here use the modern approach of effective field theory \cite{Weinberg}.
(A pedagogic introduction has been given by Kaplan\cite{Kaplan}). The energy scale relevant
for the Casimir effect is much smaller than the electron mass $m$. We can thus integrate
out the heavy electron modes from the full Lagrangian and thus obtain a low-energy effective
Lagrangian for the interacting photon field. The interactions are due to coupling of the photon
to virtual electron-positron pairs in the vacuum and they become local in the low-energy limit. To
lowest order in $\alpha$ the first effective coupling is just the Uehling correction\cite{IZ} which
modifies the photon propagator at short distances. It gives rise to the effective photon Lagrangian
\beq
     {\cal L}_{eff} = -{1\over 4}F_{\mu\nu}^2 +
                          {\alpha\over 60\pi m^2}F_{\mu\nu}\Box F^{\mu\nu}          \label{LU}
\eeq
where $F_{\mu\nu} = \del_\mu A_\nu - \del_\nu A_\mu$ is the electromagnetic field strength. If this
new interaction contributes, we see that it will modify the Casimir energy density by a term
proportional with $\alpha/m^2L^6$. Barring some extra factors of $m$ from the evaluation of
the corresponding Feynman diagram or from the integration over the volume between the plates, we see
that we are not able to reproduce the result of Bordag {\it et al} \cite{Bordag}. However,
the corresponding Feynman diagram which is a photon loop with the
Uehling correction in the propagator, is in fact zero after regularization. This is most easily
seen by using the equation of motion $\Box F_{\mu\nu} = 0$ for the photon following from the leading,
Maxwell part of the Lagrangian (\ref{LU}). One might perhaps argue that the classical equation of
motion only applies to free, i.e. on-shell photons and not to virtual particles inside a Feynman
diagram. But a more careful analysis\cite{KR} shows that the Uehling interaction does not contribute
to the vacuum energy.

The lowest order correction must thus arise from higher order interactions, i.e. of dimensions eight
or more. Among these we first have the next order Uehling term proportional with
$F_{\mu\nu}\Box^2 F^{\mu\nu}$. But this will again not contribute for the same reasons as above. But
to the same order we also have the Euler-Heisenberg interaction\cite{EH}
\beq
    {\cal L}_{EH} = {2\alpha^2\over 45 m^4}\left[({\bf E}^2 - {\bf B}^2)^2
                  + 7 ({\bf E}\cdot{\bf B})^2\right]                                  \label{LEH}
\eeq
which describes the interaction of four photons via coupling to a virtual electron loop in the vacuum.
It has previously been used by Scharnhorst and Barton to investigate the propagation of light in the
same geometry of two parallel plates\cite{SB}. For dimensional reasons it will thus give an
additional Casimir force which varies with the plate separation like $1/L^8$.

In lowest order perturbation theory the correction to the vacuum energy density is given by the
expectation value $\ex{{\cal L}_{EH}}$. For its evaluation we need the correlators $\ex{{\bf E}^2}$
and $\ex{{\bf B}^2}$ which give the fluctuations of the free electric and magnetic fields between
the plates. The modes must satisfy the metallic boundary conditions ${\bf n}\wedge {\bf E} =
{\bf n}\cdot {\bf B} = 0$ at the plates and have previously been obtained by L\"utken and
Ravndal by quantizing the field in the Coulomb gauge and expanding it in electromagnetic
multipoles\cite{LR}. They can also be obtained
from the coincidence limit of the photon propagator constructed by Bordag {\it et al} \cite{Bordag}
in the Lorentz gauge\cite{KR}. The results can be summarized as
\beq
     \ex{{\bf E}^2} = -{\pi^2\over 16L^4}\left({1\over 45} - F(\theta)\right), \hspace{10mm}
     \ex{{\bf B}^2} = -{\pi^2\over 16L^4}\left({1\over 45} + F(\theta)\right)   \label{plates}
\eeq
where $\theta = z\pi/L$ is the scaled distance between the plates and the function
\beq
     F(\theta) = -{1\over 2}{d^3\over d\theta^3}\cot{\theta}
               = {3\over\sin^4{\theta}} - {2\over\sin^2{\theta}}                 \label{F}
\eeq
gives the position dependence of the fluctuations. However, when one calculates the energy density
${\cal E}  = {1\over 2}(\ex{{\bf E}^2} + \ex{{\bf B}^2})$, it cancels out and gives ${\cal E} =
-\pi^2/720L^4$ which is the standard result for the Casimir energy per unit volume\cite{Casimir}.
When one gets very near one of the plates, the fluctuations become large. Near the left plate at
$z=0$ they increase like
\beq
     \ex{{\bf E}^2} =  {3\over 16 \pi^2 z^4},                            \hspace{10mm}
     \ex{{\bf B}^2} = - {3\over 16 \pi^2 z^4}                             \label{plane}
\eeq
This represents also the fluctuations near any curved surface since it will always look
plane when one gets near enough\cite{DC}.

With the free field electromagnetic correlators established between the plates, it is now
straightforward to calculate the lowest order correction from the Euler-Heisenberg interaction
(\ref{LEH}). In terms of Feynman diagrams, it is simply given by the contribution from the product of
two photon loops,
\beq
    \Delta{\cal E} = - {\alpha^2 \pi^4\over 2^73^35m^4L^8}
                 \left[{11\over 225} + 9F^2(\theta)\right]                        \label{deltau}
\eeq
This additional term in the Casimir energy density is now seen to diverge when one gets near one of
the plates. The divergence is directly related to the the corresponding divergence in the correlators.
We are now faced with the problem that when calculating the resulting correction to the Casimir
force, we need the total vacuum energy between the plates which is obviously also divergent.

This is
physically meaningless, but not really a new situation. For instance, when one calculates the
regularized vacuum energy density for free photons inside a spherical cavity\cite{OR}, it diverges
near the surface as in (\ref{plane}) and the integrated energy is thus also divergent. However,
if one instead calculates the total energy which afterwards is regularized, one obtains
a finite result\cite{Boyer}\cite{Bender}\cite{Milton}. That the processes of regularization and
integration do not commute in general for these kinds of problems have been discussed by Deutsch and
Candelas\cite{DC}\cite{Candelas}.

In our case with interacting fields we can obtain the total unregularized energy by integrating
the unregularized energy density. The resulting expression can then be regularized and we have a
finite, total Casimir energy. Since the calculation is somewhat cumbersome, we can illustrate it by
neglecting the uninteresting degrees of freedom parallel to the plates and consider instead the simpler
theory of a massless field in 1+1 dimensions with Lagrangian ${\cal L} = (1/2)[(\del_t\phi)^2
- (\del_z\phi)^2]$. Here $\del_t\phi$ can be taken to be the electric field and $\del_z\phi$ to be the
magnetic field. We can thus
continue to talk about the electric ${\cal E}_E = (1/2)\ex{(\del_t\phi)^2}$ and magnetic ${\cal E}_B =
(1/2)\ex{(\del_t\phi)^2}$ contributions to the vacuum energy density.
Imposing the Dirichlet condition that the field vanishes at the boundaries, the normalized
eigenmodes of the field are
\beq
       \phi_n(z) = \sqrt{2\over L}\sin{(\omega_n z)}     \label{modes}
\eeq
where $\omega_n = \pi n/L$ with $n= 1,2,3,\ldots{}$. The total vacuum energy is therefore
\beq
     E_0 = {\pi\over 2L}\sum_{n=1}^\infty n = -{\pi\over 24 L}            \label{E0}
\eeq
when we use zeta-function regularization where $\zeta(-1) = -1/12$. Assuming a constant vacuum
energy density, it is therefore $u = -\pi/24 L^2$.

Let us now consider the partial contributions to the energy. The part  coming from
fluctuations in the electric field is given by the divergent sum
\beq
       {\cal E}_E &=& {1\over 2L}\sum_{n=1}^\infty \omega_n \sin^2{(\omega_n z)}
            = {1\over 4L}\sum_{n=1}^\infty \omega_n (1 - \cos{2\omega_n z})       \label{uE1}
\eeq
If we integrate over the 1-dimensional volume to obtain the total Casimir energy, the first term
gives  $-{\pi/ 48L}$  after zeta-function regularization. But in the last position-dependent
part, we see that each term gives zero because of the imposed boundary conditions. We are
thus left with a finite contribution to the total energy which is exactly one half of the
full energy. The other half comes from the magnetic contribution.

Let us now instead derive a finite result for the energy density. It can be written as
\beq
       {\cal E}_E  = -{\pi\over 48L^2} - {\pi\over 8L^2}{d\over d\theta}S(\theta)      \label{uE2}
\eeq
where
\beq
    S(\theta) = \sum_{n=1}^\infty \sin{2\theta n}                               \label{S}
\eeq
is a divergent sum. Using again zeta-function regularization, it becomes $S_{reg}(\theta) =
{1\over 2} \cot\theta$ and we have the regularized energy density
\beq
       {\cal E}_E^{reg} = -{\pi\over 16L^2}\left({1\over 3} - {1\over \sin^2\theta}\right) \label{u_E}
\eeq
This corresponds to the result in (\ref{plates}) giving the fluctuations in the electric field between
plates. The last term gives the dependence on the position in the volume. It is positive definite and
diverges when we approach the boundaries of the volume, i.e. the end points of the interval. As a
consequence the corresponding integrated energy is also infinite. This is in sharp contrast to the
result above where the integrated contribution from the position-dependent term gave zero.

The underlying problem is obviously the behaviour of the fluctuations near the boundaries where
$\sin\theta = 0$. Then the function (\ref{S}) is zero while $S_{reg}$ is infinite. In order to better
understand this apparent paradox, we can use a more physical regularization method based upon an
exponential cutoff instead of the previous more mathematical approach using the analytical
continuation of the zeta-function. We thus define the sum (\ref{S}) by the limit $S(\theta) =
\lim_{\e\ra 0}S(\e,\theta)$ of the convergent sum
\beq
   S(\e,\theta) = \sum_{n=1}^\infty e^{-\e n}\sin{2\theta n}
                = {e^{-\e}\sin 2\theta\over 1 - 2e^{-\e}\cos 2\theta + e^{-2\e}}
\eeq
Since it is the derivative of this sum which appears in the energy density (\ref{uE2}), it is seen
not to contribute to the integrated energy even with a non-zero cutoff. This is consistent
with what we found using zeta-function regularization. However, the regularized energy density is
found in the limit $\e\ra 0$ from
\beq
   S(\e,\theta) = {1\over 2}\cot\theta - {1\over 8}\;{\cos\theta\over\sin^3\theta}\;\e^2
                + {\cal O}(\e^4)                                                      \label{expand}
\eeq
As long as $\sin\theta >0$, i.e. inside the boundaries, we can take the $\e = 0$ and we recover
the zeta-function result. But
when we get sufficiently close to the boundaries, the expansion breaks down and we have in general
a  cutoff-dependent result. This should not come as a complete surprise since the energy density
on the boundaries must be strongly dependent on the microscopic and physical properties of the material
in the boundaries. But the welcome result is that the integrated energy is independent of the cutoff
and the regularization method.

One can emulate the Euler-Heisenberg interaction in our one-dimensional system by considering the
Lagrangian
\beq
     {\cal L} = {1\over 2}(\del_\mu\phi)^2 + {\alpha\over m^2}(\del_\mu\phi)^4    \label{leh}
\eeq
Here $\alpha$ is some small dimensionless constant and $m$ a heavy mass. The Lagrangian is
invariant under the field transformation $\phi \ra \phi + const$ and thus describes massless
particles. Treating the interaction
in lowest order perturbation theory, one easily finds the resulting vacuum energy density
after regularization to be
\beq
    {\cal E} = - {\pi\over 24L^2} -  {\alpha\pi^2\over 8m^2L^4}\left({1\over 18}
        + {1\over\sin^4\theta}\right)                                            \label{uu}
\eeq
Now it diverges near the endpoints where $\sin\theta \ra 0$ and the integrated Casimir energy is
infinite. However, if we instead first integrate the energy density and then regularize as above,
we get the finite result
\beq
    E_0 = - {\pi\over 24L} -  {\alpha\pi^2\over 144m^2L^3}                        \label{E1}
\eeq
The potentially divergent part now vanishes since  $\zeta(-2) = 0$. We see that the
total Casimir energy is furnished by just the constant part of energy density (\ref{uu}). Again
there is no contribution from the position-dependent terms.

With this approach we can now calculate a finite, total Casimir energy for the photons between two
plates and interacting via the Euler-Heisenberg interaction (\ref{LEH}). Zeta-function regularization
must now be combined with dimensional regularization for the transverse degrees of freedom. As a
result we find again that the position-dependent part of the energy density (\ref{deltau}) does not
contribute to the integrated energy which now becomes\cite{KR}
\beq
    E_0 = - {\pi^2\over 720L^3}
          - {11\alpha^2\pi^4\over 2^7  3^5  5^3 m^4L^7}                              \label{E3}
\eeq
Since the energy density between the plates can thus effectively be taken to be constant in
calculating the total energy, it could also have been obtained more directly by requiring
instead the field to be periodic between the
plates with period $\beta = 2L$. This will then give a constant energy density ${\cal E} = E_0/L$
also for the interacting field. Letting $L \ra 1/2T$ in the expression for ${\cal E}$ we thus have
the free energy density for an interacting photon gas in thermal equilibrium at temperature $T$.
The result is in agreement with a previous calculation by Barton using a semi-classical
method\cite{Barton2}.
 
We want to thank Jens Andersen, Cliff Burgess and Paolo Di Vecchia for useful discussions and
clarifying comments.

\end{document}